\documentstyle[aps,preprint,epsfig,tighten]{revtex}

\newcommand{\br}{\mbox{Br}\,}
\newcommand{\tr}{\mbox{Tr}\,}
\newcommand{\kmm}{K_L\rightarrow\mu^+\mu^-}
\newcommand{\kee}{K_L\rightarrow e^+ e^-}
\newcommand{\emm}{\eta\rightarrow\mu^+\mu^-}
\newcommand{\epmm}{\eta'\rightarrow\mu^+\mu^-}
\newcommand{\eee}{\eta\rightarrow e^+ e^-}
\newcommand{\epee}{\eta'\rightarrow e^+ e^-}
\newcommand{\pee}{\pi^0\rightarrow e^+ e^-}

\newcommand{\mesgg}{P\rightarrow\gamma\gamma}
\newcommand{\mesll}{P\rightarrow l^+l^-}
\newcommand{\epll}{\eta'\rightarrow l^+l^-}

\newcommand{\kll}{K_L\rightarrow l^+l^-}
\newcommand{\kgg}{K_L\rightarrow\gamma\gamma}

\newcommand{\diag}{\mbox{diag}}
\newcommand{\gev}{\mbox{GeV}}
\newcommand{\rms}{\rm\scriptsize}

\preprint{
\begin{tabular}{r} FTUV/98$-$3 \\ IFIC/98$-$3
\end{tabular}
}

\begin{document}

\title{Long--distance contributions to the $K_L\rightarrow\mu^+\mu^-$
decay width}
\author{D.\ G\'omez Dumm and A.\ Pich}
\address{Departament de F\'{\i}sica Te\`orica, IFIC, 
CSIC -- Universitat de Val\`encia \\
Dr.\ Moliner 50, E-46100 Burjassot (Val\`encia), Spain}

\maketitle

\thispagestyle{empty}

\begin{abstract}
The dispersive two--photon contribution to the $\kmm$ decay amplitude
is analyzed,
using chiral perturbation theory techniques and large--$N_C$
considerations.
A consistent description of the decays
$\pee$, $\emm$ and $\kmm$ is obtained.
As a byproduct, one predicts
$\br (\eee) = (5.8 \pm 0.2)\times 10^{-9}$ and
$\br (\kee) = (9.0 \pm 0.4)\times 10^{-12}$.
\end{abstract}


\vspace{2cm}

The rare decay $\kmm$ has deserved a significant theoretical interest
during the last three decades. It represents a
potentially important channel to study the weak interaction within
the Standard Model (SM), as well as possible effects of new physics,
mainly in connection with flavour--changing neutral currents and CP
violation.

This decay proceeds through two distinct mechanisms:
a long--distance contribution from the $2\gamma$ intermediate state
and a short--distance part, which in the SM arises from one--loop
diagrams ({\it $W$ boxes, $Z$ penguins}) involving the weak gauge bosons.
Since the short--distance amplitude is sensitive to the presence of a 
virtual top quark, it could be used to improve our present knowledge
on the quark--mixing factor $V_{td}$; moreover, it offers a
window into new--physics phenomena. 
This possibility has renewed
the interest in the study of the $\kmm$ process in the last
years.

The short--distance SM amplitude is well--known \cite{GL:74}.
Including QCD corrections at the next-to-leading logarithm order
\cite{BB:94}, it implies \cite{BF:97}:
\begin{equation}
\label{eq:sd}
  \br(\kmm)_{\mbox{\rms SD}} = 0.9 \times 10^{-9} \; (\rho_0-\bar\rho)^2 \;
  \left({\overline{m}_t(m_t)\over 170\;\gev}\right)^{3.1}\;
  \left({\left| V_{cb}\right|\over 0.040}\right)^4
	\, , 
\end{equation}
where $\rho_0 \approx 1.2$ and $\bar\rho\equiv\rho\, (1-\lambda^2/2)$,
with $\rho$ and $\lambda$ the usual quark--mixing parameters, 
in the Wolfenstein parameterization \cite{WO:83}.
The deviation of $\rho_0$ from 1 is due to the charm contribution.
Using the presently allowed ranges for $m_t$ and the quark--mixing
factors, one gets \cite{BF:97}
$\br(\kmm)_{\mbox{\rms SD}} = (1.2\pm 0.6) \times 10^{-9}$.
If this number is compared with the measured rate \cite{Ak:95}
\begin{equation}
\label{eq:exp}
  \br(\kmm ) = (7.2\pm 0.5) \times 10^{-9} \, ,
\end{equation}
it is seen that the decay process is strongly dominated by the
long--distance amplitude.

Clearly, in order to extract useful information about the short--distance 
dynamics it is first necessary to have an accurate (and reliable)
determination of the
$K_L\rightarrow\gamma^\ast\gamma^\ast\rightarrow\mu^+\mu^-$
contribution.

It is convenient to consider the normalized ratios
\begin{equation}
\label{eq:R_def}
  R(\mesll ) \equiv
  {\br(\mesll)\over\br(\mesgg)} = 2\beta\,
  \left({\alpha\over\pi}{m_l\over M_P}\right)^2\;
  \left| F(\mesll )\right|^2 \, ,
\end{equation}
where
$\beta\equiv\sqrt{1-4m_l^2/M_P^2}$.
The on--shell $2\gamma$ intermediate state generates
the absorptive contribution \cite{MRS:70}
\begin{equation}
\label{eq:ImR}
	\mbox{Im}\ [F(\mesll )] = {\pi\over 2\beta}\, 
	\ln{\left({1-\beta\over 1+\beta}\right)} \, .
\end{equation}
Using the measured branching ratio \cite{PDG:96},
$\br (\kgg) = (5.92\pm 0.15)\times 10^{-4}$, this implies
the so-called {\it unitarity bound}:
\begin{equation}
\label{eq:abs}
  \br (\kmm ) \geq \br (\kmm )_{\mbox{\rms Abs}} = 
  (7.07\pm 0.18)\times 10^{-9} \, .
\end{equation}
Comparing this result with the experimental value in Eq.~(\ref{eq:exp}),
we see that $\br (\kmm )$ is almost saturated by the absorptive 
contribution.   

One immediate question is whether 
the small room left for the dispersive contribution,
$\br (\kmm )_{\mbox{\rms Dis}} = (0.1\pm 0.5)\times 10^{-9}$,
can be understood dynamically.
Naively, one would expect a larger value just from the
intermediate $2\gamma$ mechanism.
This has motivated some recent
speculations \cite{EKP:97} about a possible
cancellation between the long-- and short--distance dispersive
amplitudes, which could allow for additional new--physics
contributions at short distances.

The obvious theoretical framework to perform a well-defined analysis of
the long--distance part is chiral perturbation theory (ChPT).
Unfortunately, the chiral symmetry constraints are not powerful enough to
make an accurate determination of the dispersive contribution
\cite{EP:91,DIP:97,VA:97}.
The problem can be easily understood by looking at the $\kgg$ amplitude,
\begin{equation}
\label{eq:Kgg}
 A(\kgg ) = c(q_1^2,q_2^2)\; \varepsilon^{\mu\nu\rho\sigma}\,
 \epsilon_{1\mu} \epsilon_{2\nu} q_{1\rho} q_{2\sigma} \, ,
\end{equation}
which, at lowest--order in momenta, proceeds through the chain
$K_L\to \pi^0,\eta,\eta'\to 2\gamma$.
The lowest--order ---$O(p^4)$--- 
chiral prediction, can only generate a constant
form factor $c(q_1^2,q_2^2)$; it thus corresponds to the decay into
on-shell photons ($q_1^2=q_2^2=0$) \cite{ENP:92}:
\begin{equation}
\label{eq:c_Kgg}
 c(0,0) = {2 G_8 \alpha f_\pi  \over\pi (1 - r_\pi^2)}\,
 c_{\mbox{\rms red}} \, ,
\end{equation}
$$
c_{\mbox{\rms red}} = 1 - {(1 - r_\pi^2)\over 3(r_\eta^2-1)} 
  (c_\theta -2\sqrt{2} s_\theta) (c_\theta +2\sqrt{2}\rho_n s_\theta)
 + {(1 - r_\pi^2)\over 3(r_{\eta'}^2-1)} 
  (2\sqrt{2} c_\theta + s_\theta) (2\sqrt{2} \rho_n c_\theta - s_\theta)
  \, ,
$$
where
$r_P^2\equiv M_P^2/M_{K_L}^2$, $c_\theta \equiv\cos{\theta_P}$ and
$s_\theta \equiv\sin{\theta_P}$, with $\theta_P\approx - 20^\circ$
the $\eta$--$\eta'$ mixing angle.
The global parameter
$G_8\equiv 2^{-1/2} G_F V^{\phantom{*}}_{ud} V_{us}^*\, g_8$
characterizes \cite{EPR:88}
the strength of the weak $\Delta S=1$ transition
$K_L\to \pi^0,\eta,\eta'$.

In Eq. (\ref{eq:c_Kgg}) we have factored out the contribution of the pion
pole, which normalizes the dimensionless reduced amplitude
$c_{\mbox{\rms red}}$. The second and third terms in  $c_{\mbox{\rms red}}$
correspond to the $\eta$ and $\eta'$
contributions respectively. Nonet symmetry (which is exact in the
large--$N_C$ limit) has been assumed in the electromagnetic
$2\gamma$ vertices; this is known to provide a quite good description
of the anomalous $P\to 2\gamma$ decays ($P=\pi^0, \eta, \eta'$).
Possible deviations of nonet symmetry in the non-leptonic weak vertex are
parameterized through $\rho_n\not=1$.

 In the standard $SU(3)_L\otimes SU(3)_R$ ChPT, the $\eta'$ contribution
is absent  and $\theta_P=0$; therefore,
$c_{\mbox{\rms red}}\propto (3 M_\eta^2 + M_\pi^2 - 4 M_K^2)$, which vanishes
owing to the Gell-Mann--Okubo mass relation.
The physical $\kgg$ amplitude is then a higher--order ---$O(p^6)$--- effect
in the chiral counting, which makes difficult to perform a reliable
calculation.

 The situation is very different if one uses instead a
$U(3)_L\otimes U(3)_R$ effective theory \cite{HLPT:97}, 
including the singlet $\eta_1$ field.
The large mass of the $\eta'$ originates in the $U(1)_A$
anomaly which, although formally of $O(1/N_C)$, is numerically
important.
Thus, it makes sense to perform a combined chiral expansion \cite{LE:96}
in powers
of momenta and $1/N_C$, around the nonet--symmetry limit, but keeping
the anomaly contribution (i.e. the $\eta_1$ mass) together with
the lowest--order term.
In fact, the usual successful description of the $\eta/\eta'\to 2\gamma$
decays \cite{BBC:88}
corresponds to the lowest--order contribution within this
framework, plus some amount of symmetry breaking through
$f_\eta\not= f_{\eta'}\not=f_\pi$.
The mixing between the $\eta_8$ and $\eta_1$ states
provides a large enhancement of
the  $\eta\to 2\gamma$ amplitude, which is clearly
needed to understand the data.
In the standard $SU(3)_L \otimes SU(3)_R$ framework, the $\eta'$ is
integrated out and its effects are hidden in higher--order local
couplings \cite{EGPR:89}; since the $\eta_1$ and $\eta_8$
fields share the same isospin and charge, the singlet pseudoscalar does
affect the $\eta$ dynamics in a significant way, which is reflected in
the presence of important higher--order corrections~\cite{PI:90}.
These corrections are more efficiently taken into account within the
$U(3)_L\otimes U(3)_R$ framework\footnote{The contributions of the singlet
pseudoscalar are
particularly important in radiative transitions, owing to the presence of
the $\eta'$ exchange pole. A different situation occurs in the decays
$\eta\rightarrow 3\pi$, where the $O(p^4)$ corrections induced by
$\eta$--$\eta'$ mixing are related to the pseudoscalar mass
spectrum, and cancel to some extent with other contributions associated
with the exchange of scalar particles~\cite{LE2:96}.}.

  Taking $s_\theta = -1/3$ ($\theta \approx -19.5^\circ$), the $\eta$--pole
contribution in Eq.~(\ref{eq:c_Kgg}) is proportional to $(1-\rho_n)$ and
vanishes in the nonet--symmetry limit; the large and positive $\eta'$
contribution results then in $c_{\mbox{\rms red}}=1.80$ for $\rho_n=1$.
With $0\leq\rho_n\leq 1$, the $\eta$ and $\eta'$ contributions interfere
destructively and $c_{\mbox{\rms red}}$ is dominated by the pion pole. One
would get $c_{\mbox{\rms red}}\simeq 1$ for $\rho_n\simeq 3/4$.

The measured $\kgg$ decay rate \cite{PDG:96} corresponds to
$|c(0,0)| = (3.51\pm 0.05)\times 10^{-9}\;\gev^{-1}$.
With $|G_8| = 9.1\times 10^{-6}\;\gev^{-2}$, obtained from the $O(p^2)$
analysis of $K\to 2\pi$  \cite{EPR:88}, this implies
$c_{\mbox{\rms red}}^{\rms exp} = (0.84\pm 0.11)$. However,
the fitted value of $|G_8|$ gets reduced by about a 30\% when
$O(p^4)$ corrections to the $K\to 2\pi$ amplitudes are taken into account
\cite{KMW:91}. 
This sizeable shift results mainly from the constructive $\pi\pi$
rescattering contribution, which is
obviously absent in $\kgg$. Thus, we should rather use the corrected
(smaller) $|G_8|$ determination, which leads to
$c_{\mbox{\rms red}}^{\rms exp} = (1.19\pm 0.16)$.

Leaving aside numerical details, we can safely conclude that the
physical $\kgg$ amplitude, with on-shell photons, is indeed dominated
by the pion pole ($c_{\mbox{\rms red}}\sim 1$). 
Although the exact numerical prediction is sensitive to several small
corrections \cite{DHL:86} ($\rho_n\not= 1$, $f_\pi\not= f_\eta\not= f_{\eta'}$, 
$s_\theta\not=-1/3$) and therefore is quite uncertain,
the needed cancellation between
the $\eta$ and $\eta'$ contributions arises in a natural way and can be
fitted easily with a reasonable choice of symmetry--breaking
parameters.

The description of the $\kgg$ transition with off-shell photons is
a priori more complicated because the $q_{1,2}^2$ dependence of the
form factor originates from higher--order terms in the chiral lagrangian.
This is the reason why only model--dependent estimates of the
dispersive $\kll$ transition amplitude have been obtained so far.
At lowest--order in momenta, \ $c(q_1^2,q_2^2) = c(0,0)\, $; thus, the 
(divergent) photon
loop can be explicitly calculated up to a global normalization, which
is determined by the known absorptive contribution (i.e. by the
experimental value of $c(0,0)$). The model--dependence appears in
the local contributions from direct $K_L l^+ l^-$ terms
in the chiral lagrangian 
\cite{EP:91,VA:97} (allowed by symmetry considerations),
which reabsorb the loop divergence.

It would be useful to have a reliable determination in some
symmetry limit. The large--$N_C$ description of
$K_L\rightarrow\gamma^\ast\gamma^\ast$ provides such
a possibility. At leading order, this process occurs through the
$\pi^0,\eta,\eta'$ poles, as represented in Fig.\ 1. Therefore, the
problematic electromagnetic loop in Fig.\ 1(a) is actually the same
governing the decays $\pee$ and $\emm$, and the unknown local contribution
(Fig.\ 1(b)) can be fixed from the measured rates for these transitions
\cite{SLW:92,ABM:93}. In fact, the same combination of local chiral
couplings shows up in both decays \cite{SLW:92}, leading to a relation that
is well satisfied by the data.

Although the $\epll$ transition introduces additional chiral couplings,
they are suppressed by at least one more power of $1/N_C$.
Thus, in the large--$N_C$ limit the different electromagnetic
$\mesll$ decays get related through the same counterterms \cite{SLW:92}:
\begin{equation}
\label{eq:counterterm}
 {\cal L}_{\mbox{\rms c.t.}} \, = \,
  {3 i \alpha^2\over 32\pi^2}\; (\bar l\gamma^\mu\gamma_5 l)\, \left\{
  \chi_1\, \tr\left(
  Q^2 \{U^\dagger ,\partial_\mu U\} \right) 
  + \chi_2\, \tr\left(
  Q U^\dagger Q \partial_\mu U - Q \partial_\mu U^\dagger Q U\right) 
  \right\}\, ,
\end{equation}
where $Q\equiv\diag (2/3,-1/3,-1/3)$ is the quark electromagnetic
charge matrix and $U\equiv \exp{\left(i\sqrt{2} \Phi /f\right)}$ the
usual $U(3)_L\otimes U(3)_R$
matrix describing the pseudoscalar nonet.

Nonet symmetry should provide a good estimate of the ratio
$R(\kll)$. Since $\kgg$
is dominated by the pion pole, we can reasonably expect 
that symmetry--breaking corrections would play a rather small role.
In any case, this symmetry limit allows us to investigate whether
the tiny dispersive contribution allowed by the data is what should
be expected from the $2\gamma$ intermediate state.

  In this limit, all $R(\mesll)$ ratios are
governed by the same dispersive amplitude\footnote{Notice that our result
differs slightly from those quoted in Refs.~\cite{ABM:93} and \cite{DE:86}.
We agree with the numerical expression given in
Ref.~\protect\cite{SLW:92}.}:
\begin{equation}
\label{eq:disp}
\mbox{Re\ }[F(\mesll )] \, = \, 
{1\over 4\beta} \ln^2{\left({1-\beta\over 1+\beta}\right)}
+ {1\over \beta} Li_2\left({\beta-1\over\beta +1}\right)
+ {\pi^2\over 12\beta} + 3 \ln{\left({m_l\over\mu }\right)} 
+ \chi(\mu)
\, ,
\end{equation}
where
$\chi(\mu) \equiv -\left(\chi_1^r(\mu) + \chi_2^r(\mu) + 14 \right)/4$
is the relevant local contribution, with
$\chi^r_i(\mu)$ (i=1,2) the corresponding
chiral couplings renormalized in the $\overline{\mbox{MS}}$ scheme.
The $\mu$ dependence of the $\chi(\mu)$ and $\ln{\left(m_l/\mu\right)}$
terms compensate each other, so that the total amplitude
is $\mu$--independent.

Table \ref{tab:chi} shows the fitted values of $\chi (M_\rho)$ from
the three measured ratios $R(\pee )$, $R(\emm )$ and $R(\kmm )$.
Subtracting the known absorptive contribution, the experimental data 
provide two possible solutions for each ratio;
they correspond to a total positive (solution 1) or negative (solution 2)
dispersive amplitude. We see from the Table that the second solution from
the decay $\pee$ is clearly ruled out; owing to the smallness of the
electron mass,
the logarithmic loop contribution dominates the dispersive amplitude,
which has then a definite positive sign
(an unnaturally large and negative value of $\chi(M_\rho)$ is needed
to make it negative).
The large experimental errors do not allow to discard at this point
any of the other solutions: the remaining value from $\pee$ is consistent
with the results from the $\emm$ and $\kmm$ decays, and these are
also in agreement with each other if the same solution (either the first
or the second) is taken for both. We see that, in any case, the three
experimental ratios are well described by a common value of
$\chi (M_\rho)$. In this way, the experimentally observed small
dispersive contribution to the $\kmm$ decay rate fits perfectly well
within the large--$N_C$ description of this process.

We have not considered up to now the short--distance
contribution to the $\kmm$ decay amplitude \cite{BF:97}. This can be done
through a shift of the effective
$\chi(M_\rho)$ value\footnote{The relative sign between the short-- and
long--distance dispersive amplitudes is fixed by the known positive sign
of $g_8$ in the large--$N_C$ limit \cite{PR:96}.}:
\begin{equation}
\chi(M_\rho)_{\mbox{\rms eff}} = \chi(M_\rho) - \delta\chi_{\mbox{\rms SD}}
\, ,
\end{equation}
$$
\delta\chi_{\mbox{\rms SD}} \,\approx\, 1.7\, \left(\rho_0 -\bar\rho\right)\;
\left({\overline{m}_t(m_t)\over 170\;\gev}\right)^{1.56}\;
  \left({\left| V_{cb}\right|\over 0.040}\right)^2
  \, .
$$ 
For the allowed range $|\bar\rho|\leq 0.3$, one has
$\delta\chi_{\mbox{\rms SD}}\approx 1.8\pm 0.6$, which allows to
exclude the solution 2 for $\chi(M_\rho)$ obtained from $\emm$.
The solution 1, on the contrary, is found to be compatible with
the results from $\kmm$, and can be used to get a constraint for
$\delta\chi_{\mbox{\rms SD}}$. Indeed, taking as the best determination
\begin{equation}
\chi(M_\rho) = 5.5\, {}^{+0.8}_{-1.0}\; ,
\label{eq:chi}
\end{equation}
the first solution for $\kmm$ leads to
\begin{equation}
\delta\chi_{\mbox{\rms SD}} =
2.2\, {}^{+1.1}_{-1.3} \; ,
\label{eq:dchi}
\end{equation}
in agreement with the $\delta\chi_{\mbox{\rms SD}}$ value quoted above.
The second solution for $\kmm$ appears to be less favoured, yielding
$\delta\chi_{\mbox{\rms SD}}=3.6\pm 1.2$; this shows a discrepancy of
about $1.4\,\sigma$ with the short--distance estimate. Notice that the
precision of the result in (\ref{eq:dchi}) is still relatively low. However,
the errors could be reduced by improving the measurements of the $\emm$
and $\kmm$ branching ratios.

Finally, once the local contribution to the $\mesll$ decay amplitude has
been fixed, it is possible to obtain definite predictions for the
decays into $e^+ e^-$ pairs:
\begin{equation}
\begin{array}{ll}
\mbox{Br}(\pee) = (8.3 \pm 0.4)\times 10^{-8} \, , \quad &
\mbox{Br}(\eee) = (5.8 \pm 0.2)\times 10^{-9} \, , \\
\mbox{Br}(\kee) = (9.0 \pm 0.4)\times 10^{-12} \, .
\end{array}
\end{equation}
In the same way, the amplitudes corresponding to the $\eta'$ decays are
found to be $\mbox{Br}(\epee) = (1.5 \pm 0.1)\times 10^{-10}$ and
$\mbox{Br}(\epmm) = (2.1 \pm 0.3)\times 10^{-7}$. However, in view of the
large mass of the $\eta'$, these predictions could receive important
corrections from higher--order terms in the chiral lagrangian.

To summarize, in the nonet symmetry limit it is possible to make
a reliable determination of the ratios $R(\mesll)$, at lowest
non-trivial order in the chiral expansion. 
A consistent picture of all measured $\mesll$ decay modes is obtained,
within the SM. The present data allow
to pin down the size of the relevant chiral counterterm and
to get a constraint on the short--distance contribution
to the $\kmm$ amplitude. However, this constraint is found to
be rather weak; more precise measurements of the $\emm$ and
$\kmm$ branching ratios would be necessary in order to improve the
bounds for $\delta\chi_{\mbox{\rms SD}}$ obtained through
the SM box and penguin computations. In addition, a more detailed
investigation of the theoretical uncertainties is needed to quantify
how precisely the short--distance $\kmm$ amplitude can be inferred
from the data. Although it seems difficult to achieve a theoretical
precision good enough for making useful tests of the SM flavour--mixing
structure, it is worth to try.

\acknowledgements

We would like to thank Ll.\ Ametller, J.\ Portol\'es and J.\ Prades for
useful discussions. This work has been supported in part by 
CICYT (Spain) under grant No. AEN-96-1718. The work of
D. G. D.\ has been supported by a grant from the Commission
of the European Communities, under the TMR programme (Contract
N$^\circ$ ERBFMBICT961548).

\begin{figure}[htbp]
\begin{center}
\vspace{1cm}
\epsfig{file=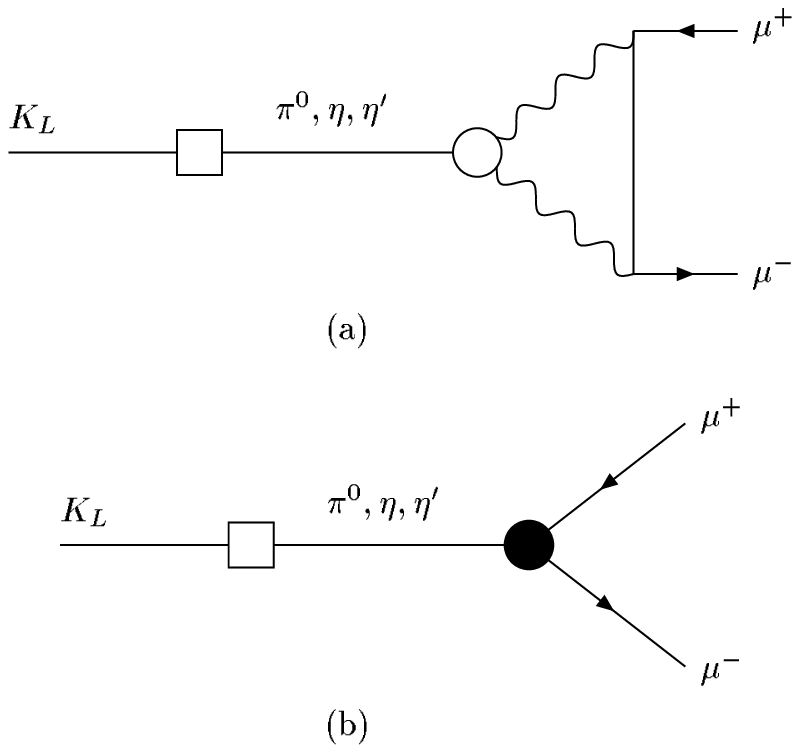}
\end{center}
\caption{(a) Photon loop and (b) associated counterterm contributions
to the $\kmm$ process.}
\end{figure}

\begin{table}
\caption{Fitted values of $\chi (M_\rho)$ from different $R(\mesll)$ ratios.
The numbers quoted for $\kmm $ refer to the difference
$\chi (M_\rho) - \delta\chi_{\mbox{\protect\rms SD}}$.}
\label{tab:chi}
\begin{tabular}{ccc}
& $\chi (M_\rho)$ [Solution 1] & $\chi (M_\rho)$ [Solution 2]
\\ \hline
$\pee $ & $4\,{}^{+4}_{-6}$ & $-24\pm 5$ \\
$\emm $ & $5.5\,{}^{+0.8}_{-1.0}$ & $-0.8\,{}^{+1.0}_{-0.8}$  \\
$\kmm $ & $3.3\,{}^{+0.9}_{-0.7}$ & $1.9\,{}^{+0.7}_{-0.9}$  \\
\end{tabular}
\end{table}

\end{document}